\documentclass[aip,rsi,reprint,footinbib]{revtex4-1}

\usepackage{graphicx}
\usepackage{epstopdf}
\usepackage{dcolumn}
\usepackage{bm}

\begin{document}


\title{A simple radionuclide-driven single-ion source}

\author{M.~Montero~D\'{i}ez} \affiliation{Physics Department, Stanford University, Stanford CA, USA}
\author{K.~Twelker} \affiliation{Physics Department, Stanford University, Stanford CA, USA}
\author{W.~Fairbank Jr.}\affiliation{Physics Department, Colorado State University, Fort Collins CO, USA}
\author{G.~Gratta} \affiliation{Physics Department, Stanford University, Stanford CA, USA}
\author{P.S.~Barbeau} \affiliation{Physics Department, Stanford University, Stanford CA, USA}
\author{K.~Barry} \affiliation{Physics Department, Stanford University, Stanford CA, USA}
\author{R.~DeVoe} \affiliation{Physics Department, Stanford University, Stanford CA, USA}
\author{M.J.~Dolinski} \affiliation{Physics Department, Stanford University, Stanford CA, USA}
\author{M.~Green} \affiliation{Physics Department, Stanford University, Stanford CA, USA}
\author{F.~LePort} \affiliation{Physics Department, Stanford University, Stanford CA, USA}
\author{A.R.~M\"{u}ller} \affiliation{Physics Department, Stanford University, Stanford CA, USA}
\author{R.~Neilson} \affiliation{Physics Department, Stanford University, Stanford CA, USA}
\author{K.~O'Sullivan}\affiliation{Physics Department, Stanford University, Stanford CA, USA}
\author{N.~Ackerman}\affiliation{SLAC National Accelerator Laboratory, Menlo Park CA, USA}
\author{B.~Aharmin}\affiliation{Physics Department, Laurentian University, Sudbury ON, Canada}
\author{M.~Auger}\affiliation{LHEP, Physikalisches Institut, University of Bern, Bern, Switzerland}
\author{C.~Benitez-Medina}\affiliation{Physics Department, Colorado State University, Fort Collins CO, USA}
\author{M.~Breidenbach}\affiliation{SLAC National Accelerator Laboratory, Menlo Park CA, USA}
\author{A.~Burenkov}\affiliation{Institute for Theoretical and Experimental Physics, Moscow, Russia}
\author{S.~Cook}\affiliation{Physics Department, Colorado State University, Fort Collins CO, USA}
\author{T.~Daniels}\affiliation{Physics Department, University of Massachusetts, Amherst MA, USA}
\author{K.~Donato}\affiliation{Physics Department, Laurentian University, Sudbury ON, Canada}
\author{J.~Farine}\affiliation{Physics Department, Laurentian University, Sudbury ON, Canada}
\author{G.~Giroux}\affiliation{LHEP, Physikalisches Institut, University of Bern, Bern, Switzerland}
\author{R.~Gornea}\affiliation{LHEP, Physikalisches Institut, University of Bern, Bern, Switzerland}
\author{K.~Graham}\affiliation{Physics Department, Carleton University, Ottawa ON, Canada}
\author{C.~Hagemann}\affiliation{Physics Department, Carleton University, Ottawa ON, Canada}
\author{C.~Hall}\affiliation{Physics Department, University of Maryland, College Park MD, USA}
\author{K.~Hall}\affiliation{Physics Department, Colorado State University, Fort Collins CO, USA}
\author{D.~Hallman}\affiliation{Physics Department, Laurentian University, Sudbury ON, Canada}
\author{C.~Hargrove}\affiliation{Physics Department, Carleton University, Ottawa ON, Canada}
\author{S.~Herrin}\affiliation{SLAC National Accelerator Laboratory, Menlo Park CA, USA}
\author{A.~Karelin}\affiliation{Institute for Theoretical and Experimental Physics, Moscow, Russia}
\author{L.J.~Kaufman}\thanks{Now at Indiana University, Bloomington IN, USA}\affiliation{Physics Department, University of Maryland, College Park MD, USA}
\author{A.~Kuchenkov}\affiliation{Institute for Theoretical and Experimental Physics, Moscow, Russia}
\author{K.~Kumar}\affiliation{Physics Department, University of Massachusetts, Amherst MA, USA}
\author{J.~Lacey}\affiliation{Physics Department, Carleton University, Ottawa ON, Canada}
\author{D.S.~Leonard}\thanks{Now at the University of Seoul, Seoul, Korea}\affiliation{Physics Department, University of Maryland, College Park MD, USA}
\author{D.~Mackay}\affiliation{SLAC National Accelerator Laboratory, Menlo Park CA, USA}
\author{R.~MacLellan}\affiliation{Dept. of Physics and Astronomy,University of Alabama, Tuscaloosa AL,USA}
\author{B.~Mong}\affiliation{Physics Department, Colorado State University, Fort Collins CO, USA}
\author{E.~Niner}\affiliation{Dept. of Physics and Astronomy,University of Alabama, Tuscaloosa AL,USA}
\author{A.~Odian}\affiliation{SLAC National Accelerator Laboratory, Menlo Park CA, USA}
\author{A.~Piepke}\affiliation{Dept. of Physics and Astronomy,University of Alabama, Tuscaloosa AL,USA}
\author{A.~Pocar}\affiliation{Physics Department, University of Massachusetts, Amherst MA, USA}
\author{C.Y.~Prescott}\affiliation{SLAC National Accelerator Laboratory, Menlo Park CA, USA}
\author{K.~Pushkin}\affiliation{Dept. of Physics and Astronomy,University of Alabama, Tuscaloosa AL,USA}
\author{E.~Rollin}\affiliation{Physics Department, Carleton University, Ottawa ON, Canada}
\author{P.C.~Rowson}\affiliation{SLAC National Accelerator Laboratory, Menlo Park CA, USA}
\author{D.~Sinclair}\affiliation{Physics Dept., Carleton University, Ottawa and TRIUMF, Vancouver, Canada}
\author{S.~Slutsky}\affiliation{Physics Department, University of Maryland, College Park MD, USA}
\author{V.~Stekhanov}\affiliation{Institute for Theoretical and Experimental Physics, Moscow, Russia}
\author{J.-L.~Vuilleumier}\affiliation{LHEP, Physikalisches Institut, University of Bern, Bern, Switzerland}
\author{U.~Wichoski}\affiliation{Physics Department, Laurentian University, Sudbury ON, Canada}
\author{J.~Wodin}\affiliation{SLAC National Accelerator Laboratory, Menlo Park CA, USA}
\author{L.~Yang}\affiliation{SLAC National Accelerator Laboratory, Menlo Park CA, USA}
\author{Y.-R.~Yen}\affiliation{Physics Department, University of Maryland, College Park MD, USA}

\date{\today}

\begin{abstract}
We describe a source capable of producing single barium ions through nuclear recoils in radioactive decay. The source is fabricated by electroplating $^{148}$Gd onto a silicon $\alpha$-particle detector and vapor depositing a layer of BaF$_2$ over it. $^{144}$Sm recoils from the alpha decay of $^{148}$Gd are used to dislodge Ba$^+$ ions from the BaF$_2$ layer and emit them in the surrounding environment. The simultaneous detection of an $\alpha$ particle in the substrate
detector allows for tagging of the nuclear decay and of the Ba$^+$ emission. The source is simple, durable, and can be manipulated and used in different environments. We discuss the fabrication process, which can be easily adapted to emit most other chemical species, and the performance of the source.
\end{abstract}

\maketitle

\section{Introduction}

It is often necessary in experimental settings to produce specific ionic species in a controlled manner. Indeed a large number of techniques to produce ions have been documented, from the early Penning sources to laser and electron-beam produced plasmas to thermal ionization~\cite{Valyi,Wolf}. These sources, however, can only operate under vacuum, and in many cases require expensive and complex infrastructure to allow for ion separation and delivery.

The work discussed here arises from the need to develop a simple source of monoatomic, singly-ionized barium that can be used in a cryogenic liquid, a gaseous environment, and in vacuum. Such a source is necessary for the R\&D efforts of the Enriched Xenon Observatory (EXO). The EXO Collaboration is planning a series of experiments designed to determine the mass of the neutrino by searching for the neutrinoless double-beta decay~\cite{doublebeta} of $^{136}$Xe. To suppress possible radioactive backgrounds, EXO plans to add to the standard low background techniques that are customary in these experiments the ability to tag the barium daughter of the double-beta decay using resonant fluorescence~\cite{M.Danilov,B.Flatt,M.Green}.   Since possible detector technologies would include xenon Time Projection Chambers (TPCs) in high pressure gas~\cite{M.Danilov} or liquid~\cite{E.Conti} phase, the R\&D work on Ba-tagging requires a method of releasing Ba$^+$ in such media.    The injection through a thin window of Ba ions produced in an accelerator would clearly be possible but it would require expensive equipment and would result in very highly ionized states, owing to the 
electron-stripping in the entrance window and in the high density medium.

The source presented here makes use of the Sm daughter of the decay
\begin{equation}
{\rm ^{148}_{64}Gd \rightarrow \ ^{144}_{62}Sm +\alpha}
\end{equation}
to dislodge a Ba$^+$ ion from a BaF$_2$ layer coated over the Gd source.   The $\alpha$
recoiling against the Sm is then detected in a semiconductor detector that is used as a substrate 
for the assembly, providing a tag for the ion emission.     $^{148}$Gd is a convenient $\alpha$
emitter because its half-life (74.6~y) provides a good compromise between high specific activity and durability.   In addition, the decay of $^{148}$Gd is directly in the ground state of $^{144}$Sm, so that no other radioactivity is produced.  Finally, Gd can be reliably electroplated in thin, uniform
layers, as will be discussed.

The Gd source is deposited onto a PIPS\textsuperscript{\textregistered} $\alpha$ 
detector~\cite{wafer}
and a special technique is used to confine the activity to a small central region.    
A thin layer of BaF$_2$ is then vapor-deposited 
over the entire detector.    The assembly is mounted in a custom metal-ceramic holder that is
suitable for ultra-clean operations and moderate vacuum bakeouts.    The ion source obtained is
inexpensive, compact and suitable for a variety of environments.    The selection of coatings other
than BaF$_2$ can provide similar sources for most elements.

\section{Source Fabrication}

$^{148}$Gd is obtained in the form of GdCl$_3$ in HCl aqueous solution.    Natural 
Gd is present as carrier in the solution in the approximate ratio~\cite{carrier}
$^{\rm nat}$Gd:$^{148}$Gd=1:6.6$\times 10^{-4}$. Gd electroplating 
works best in isopropanol (IPA) from gadolinium nitrate\cite{Moody}.   So the 
chloride is transformed into nitrate as:
\begin{equation}
 {\rm GdCl_{3(aq)}+3HNO_{3(aq)}\rightarrow Gd(NO_3)_{3(aq)}+3HCl_{(aq)}}
\end{equation}

The reaction is achieved by drying a 500$\mu$L (0.1$\mu$Ci) $^{148}$GdCl$_3$ batch on
a hot plate and adding 500$\mu$L of 1M HNO$_3$~\cite{acid}. 
The water is then evaporated again and more HNO$_3$ is added, repeating the process three times.
After the last drying cycle, the $^{148}$Gd(NO$_3$)$_3$ is dissolved in 500$\mu$L of IPA~\cite{IPA}.  25$\mu$L of this solution are calculated to have an activity of 185~Bq.   A measurement of the activity, performed by 
dissolving one 25$\mu$L batch in a liquid scintillation counter~\cite{counting}, confirms a carry-over efficiency 
of $\sim$100\%.   Since the Gd(NO$_3$)$_3$
concentration is extremely low, a drop of 0.05M HNO$_3$ is added in 10L of the IPA in order to
increase the conductivity of the plating solution to an acceptable level.

\begin{figure}
\includegraphics[angle=0,scale=0.34]{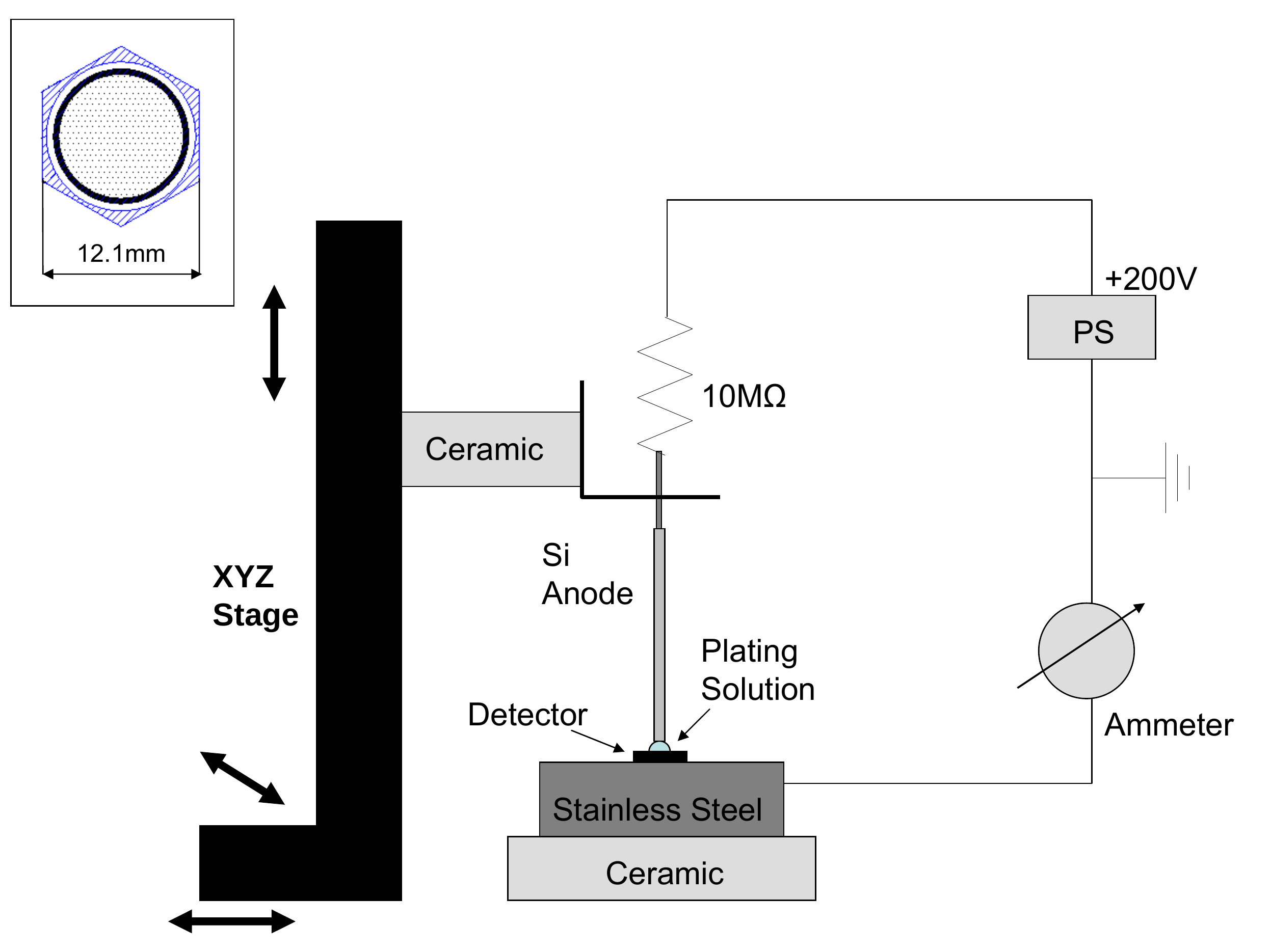}
\caption{\label{fig:plating} Schematic view of the radioactive Gd plating setup.    The micrometers
allow for an accurate and stable positioning of the anode $\simeq 0.5$~mm above the center of the 
detector being plated.    A drawing of the Canberra PIPS\textsuperscript{\textregistered} 
detector used is provided in the inset.}
\end{figure}

The use of a conventional plating cell is excluded because of the requirement to deposit the $^{148}$Gd
only in a specific and small region of the detector.     Therefore, a plating circuit is established
by mounting a PIPS\textsuperscript{\textregistered} detector on a horizontal support and lowering an anode to a position about 0.5~mm above the center 
of the detector using a micrometer, as shown in Figure~\ref{fig:plating}.    Using a calibrated
pipette a small drop of $\sim 5 \mu$L of solution is then deposited between the anode and the 
detector, which serves as the cathode in the circuit.   The drop remains in place because of 
surface tension.
A 200~V potential is then applied to the circuit via a 10~M$\Omega$ current-limiting resistor, 
establishing a current that is measured to be 18~$\mu$A.    Depending on the activity required more 5$\mu$L batches are 
added, as the solvent evaporates, for a total of 25 to 50$\mu$L.    Finally, pure IPA aliquots are 
added, again to counter evaporation.     A total plating time of 15~min was found to be adequate 
by trial and error by measuring the activity on the detector (since the charge transport in the 
cell is completely dominated by the HNO$_3$ added to the IPA and has no relation with the 
amount of Gd deposited).       Because of the non-radioactive Gd carrier, a source of 
$\simeq$200~Bq activity,
obtained from 25$\mu$L plating solution, roughly covers a surface of $8\times 10^{-2}$~cm$^2$ 
with $\sim$10 monolayers of Gd.

Initial tests using Pt anodes were found to produce large contaminations in the Gd coatings.  This
was evident by optical inspection of the plated detector.   Such contaminations are 
unacceptable because they reduce the energy of the Sm recoils, as confirmed by the analysis of 
the $\alpha$ energy spectrum from the source measured in an external surface barrier detector.    
Anodes obtained using 4~mm - wide slices of a B-doped silicon wafer (resistivity $\approx 50$
~$\Omega$m) were found to be sufficiently clean and were used for the work presented here.

\begin{figure}
\includegraphics[angle=90,scale=0.45]{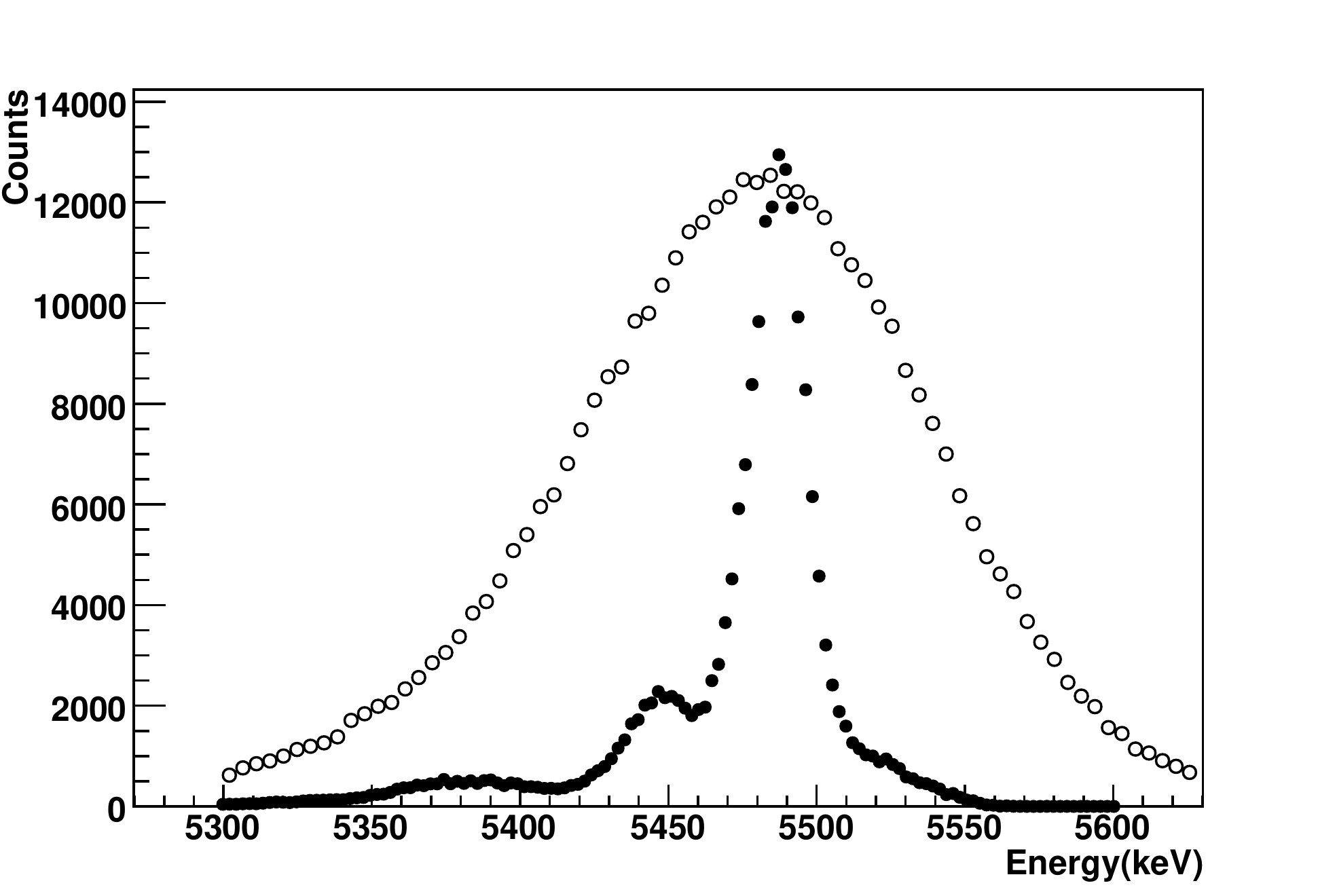}
\caption{\label{fig:alpha} $\alpha$ spectrum of an external $^{241}$Am source recorded by the
PIPS\textsuperscript{\textregistered}
detector before (full circles) and after (empty circles) the Gd plating.  The three main $\alpha$ peaks at
5485.6~keV, 5442.8~keV and 5388.2~keV are clearly distinguishable in the first data set.  In 
this case the resolution of the 5485.6~keV peak is of 22~keV FWHM.    After plating, the energy 
resolution degrades substantially and the new resolution of $\sim 130$~keV FWHM makes it 
impossible to resolve the three $^{241}$Am lines.}
\end{figure}

\begin{figure}
\includegraphics[scale=0.55]{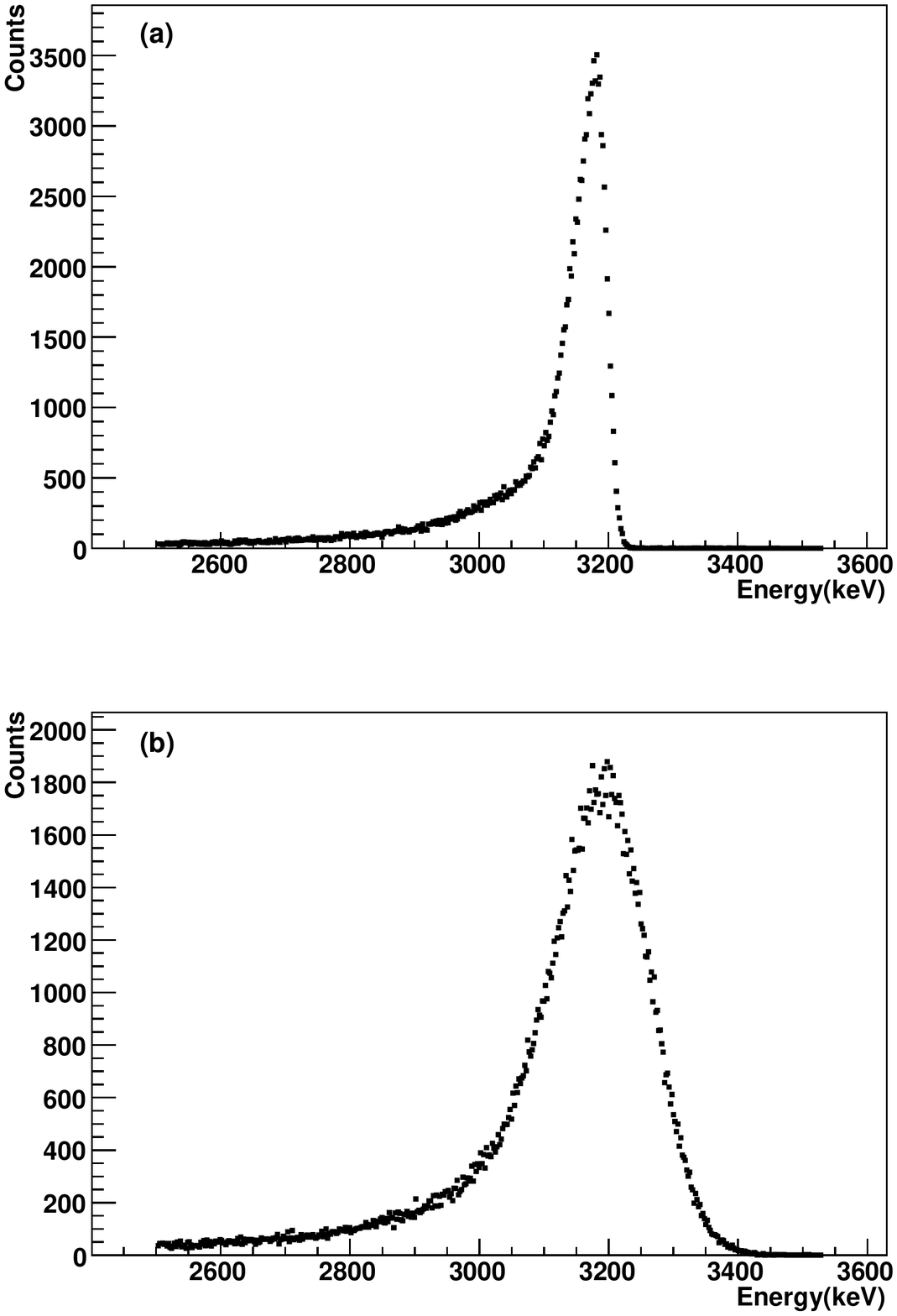}
\caption{\label{fig:gdalpha} $^{148}$Gd ``self-counted'' $\alpha$ spectra. 
Panel (a) shows the spectrum of the $^{148}$Gd layer only. The FWHM energy resolution
is $\sim$60~keV. Panel (b) shows the spectrum after 40~nm of BaF$_2$ 
are deposited over the source. The FWHM energy resolution is in this case 
$\sim$170~keV.}
\end{figure}

The $\alpha$ detector used as a substrate has surface contacts formed by shallow ion implantation and 
it is especially appropriate because of its ruggedness.    Its hexagonal shape (also shown 
in Figure~\ref{fig:plating}) is of no particular significance here.   The functionality of the 
detector after plating and the quality of the plated source are verified by $\alpha$-counting
both with the source-substrate detector (``self-counting'') and, in some cases, with an 
external surface-barrier
detector~\cite{detector} in a vacuum chamber.    In addition, an external $^{241}$Am source is
used to verify the detector performance and resolution after plating.     
A typical $^{241}$Am spectrum recorded 
by the PIPS\textsuperscript{\textregistered} detector with a 50~V bias and 2~$\mu$s shaping 
time, before and after Gd plating is 
shown in Figure~\ref{fig:alpha}.  The three peaks at 5485.6~keV, 5442.8~keV and 5388.2~keV are 
clearly visible before plating, with a 22~keV FWHM resolution for the most 
prominent peak.    This is somewhat worse than the 14~keV FWHM stated by the manufacturer.   
The spectrum after Gd plating is substantially broader ($\sim 130$~keV FWHM), presumably due 
to some deterioration of the detector and some energy loss in the Gd and other impurities 
left on the surface by the plating process.   
This degradation, however, appears to be stable in time and is not important for the simple
$\alpha$ detection required in tagging the ion emission.      The spectrum from the
single 3182.7~keV $\alpha$ emitted by the $^{148}$Gd plated on a detector shows 
(Figure~\ref{fig:gdalpha}(a)) a resolution of $\sim 60$~keV FWHM.

The activity of the plated $^{148}$Gd is measured by reading out the signal from the $\alpha$ detector, as will be described in the next section.   For typical sources this activity is $\simeq$200~Bq
and matches, within experimental uncertainties, the activity measured in the plating solution, 
implying that the plating efficiency is rather high.

A coating of the appropriate chemical species and thickness will generally produce free ions and
neutral atoms dislodged by the collision of the 89~keV  
recoiling $^{144}$Sm.    This is expected to be true for most elements, although with 
varying efficiencies, mainly due to the ratio of masses between the Sm and the ``target'' species.  
Different (in particular, heavier) $\alpha$ emitters (e.g. $^{208}$Po or $^{209}$Po) could
replace the $^{148}$Gd, although this was not attempted in our work.

\begin{figure}
\includegraphics[scale=0.47]{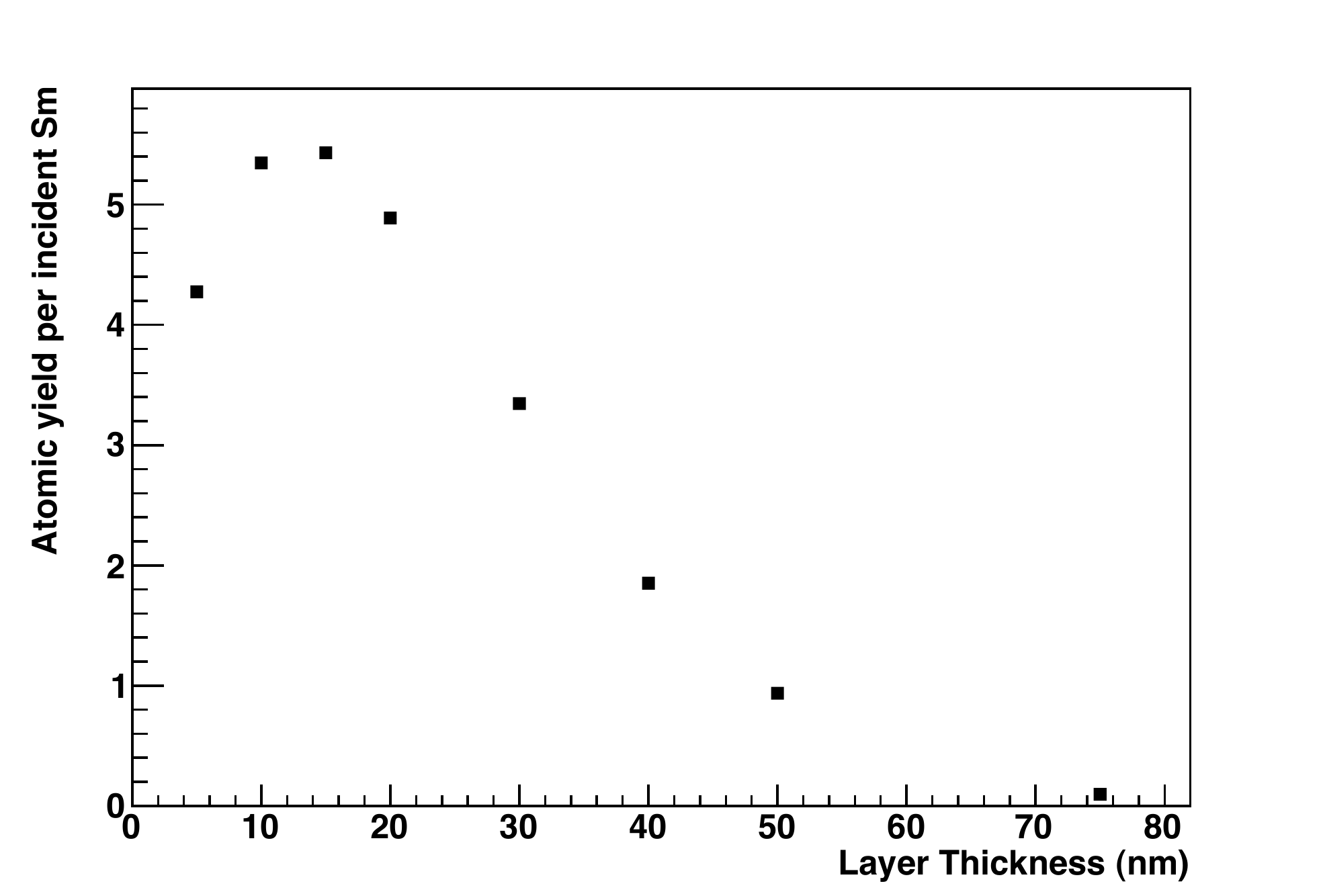}
\caption{\label{fig:thickness} Yield for Ba atoms as a function of layer thicknesses, as
calculated by a SRIM simulation.  Since SRIM does not distinguish between different 
charge states the yield from the graph includes neutral atoms, singly charged ions as well as
ions with a higher ionization state.}
\end{figure}

BaF$_2$ was chosen as the chemical species
containing Ba because of its relatively good stability in air~\cite{unhygro} and the fact that 
the other atomic species, fluorine, has a mass 
that is very different from that of Ba.   
The optimal layer thickness is estimated using the SRIM~\cite{SRIM} simulation package with full damage 
cascades simulation.   Figure~\ref{fig:thickness} shows the predicted Ba yields as 
a function of thickness.  BaF$_2$ was simulated by simply providing SRIM with the appropriate
number densities of Ba and F.  
Yields greater
than one, like those reported in Figure~\ref{fig:thickness}, are not unusual in sputtering processes.

 While the Ba yield can be larger than 1, as in the case of sputtering,
it should be kept in mind that SRIM does not distinguish between neutral atoms and ions in
different charge states, so that the yield of Ba$^+$ is not known from the simulation.  The 
energy spectrum for Ba is predicted by SRIM to be peaked at zero with a long tail 
extending above 100~eV.     15~nm was chosen as the optimal layer thickness for the BaF$_2$,
although some early sources were fabricated with thicknesses from 8~nm to 40~nm.    

As already mentioned, electroplating was chosen as the technique to deposit the $\alpha$ 
emitter because of the layer quality and the possibility of obtaining deposits that are easily
limited to a specific area.    The technology for the overcoating by the target species 
was then chosen so as not to chemically interfere with the radionuclide and, at the same time, 
provide a uniform layer over the entire front surface of the detector.    Vacuum evaporation 
is suitable for this since the energy of the ions is low (compared to sputtering) and no wet 
chemistry is involved.    In our system, tantalum boats are ohmically heated to an appropriate 
temperature and the thickness of the deposited layer is monitored by a microbalance located
next to the source being plated.    Most materials can be deposited in this way or using 
electron-beam heating for refractory materials.
As shown in Figure~\ref{fig:gdalpha}(b), the $\alpha$ energy resolution from the $^{148}$Gd 
source is further deteriorated after the BaF$_2$ coating.    While the origin of 
the deterioration is not understood, the detector is still functional and 
completely adequate to tag the decay and the subsequent emission.
In order to ensure the best possible purity for the BaF$_2$ coating, a small scintillation-grade 
BaF$_2$ crystal~\cite{crystal} was used in the 
evaporation process.

\begin{figure}
\includegraphics[scale=0.43]{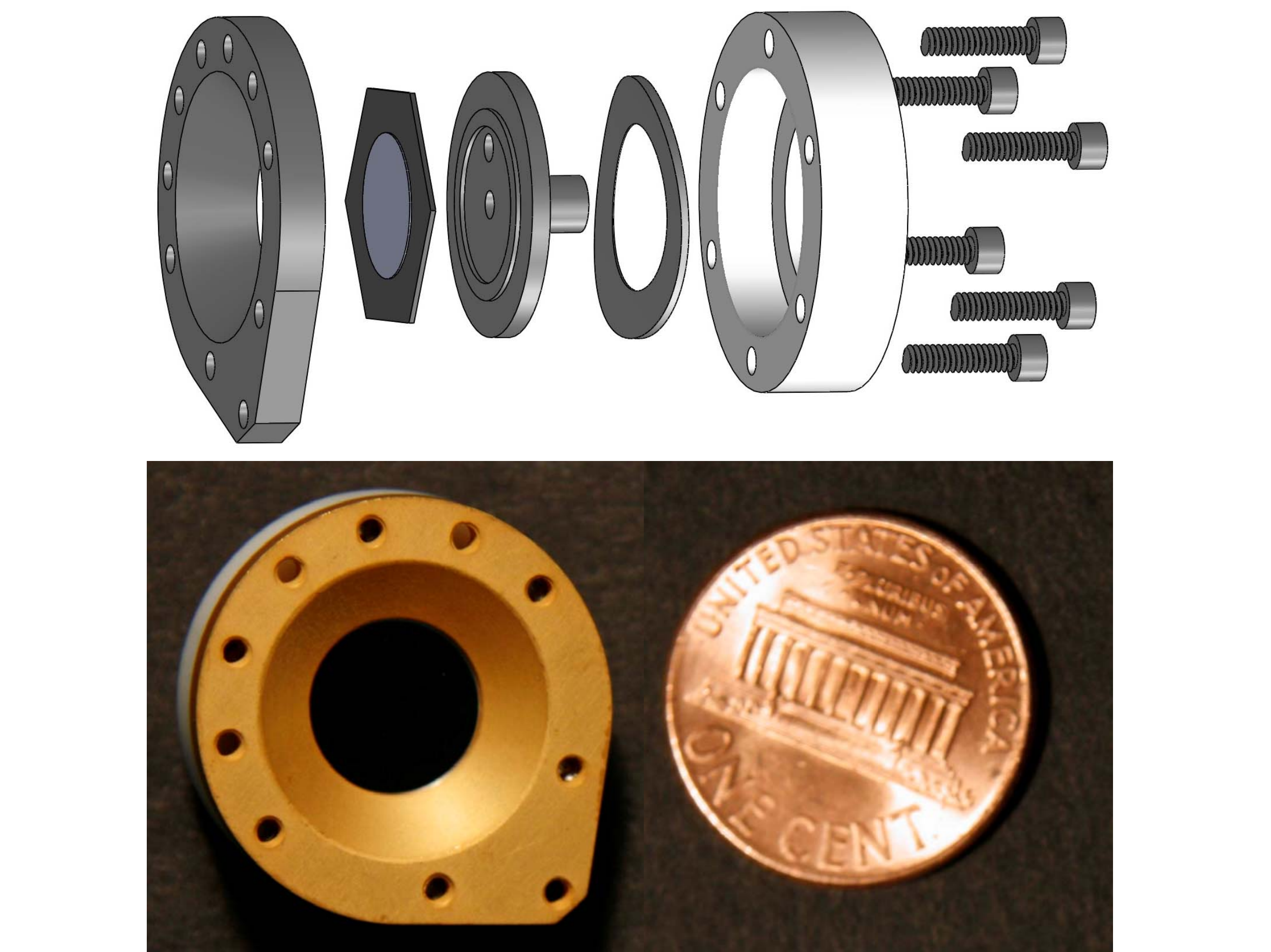}
\caption{\label{fig:source} Top: Exploded view of an ion source in its metal-ceramic UHV mount.
The ``wavy washer'' provides a uniform force on the detector to ensure proper contact.   All 
metallic parts are made out of gold-plated stainless steel.    The ring to the right of the drawing
is made out of Macor\textsuperscript{\textregistered}.   Bottom: Photograph of a completed source.}
\end{figure}

Sources prepared as described above are then mounted in the special holders shown in 
Figure~\ref{fig:source}.      All metallic parts are made of gold-plated stainless steel, and 
the insulating ring separating the two diode contacts is made out of 
Macor\textsuperscript{\textregistered}.
The elastomer used in the standard encapsulation to maintain a constant pressure on the contacts
is replaced by a gold-plated wavy washer.    The holders allow for cooldown to 77~K and are
compatible with the extreme cleanliness required for ultra-high vacuum operations as well as
for operation in liquefied noble gases (in particular liquid xenon).

\section{Source Characterization}

The ion emission from the source is analyzed with a simple time of flight (TOF) spectrometer
comprising two Einzel lenses and designed to maximize the transport efficiency of ions from 
the rather large emittance produced by the source to a channel electron 
multiplier~\cite{CEM} (CEM).     The spectrometer is designed 
with a central drift region and electrodes at the two ends that allow for the independent 
adjustment of the electric field at the surface of the source ($E_{\rm source}$,
$350~{\rm V/cm} < E_{source} < 750~{\rm V/cm}$) and the ion impact energy on the CEM 
($K$ $2400~{\rm eV} < K < 3800~{\rm eV}$).
The overall length of $\simeq 17$~cm produces a TOF for Ba$^+$ of $\sim$5~$\mu$s, depending
on the exact configuration of potentials.
The ``start'' signal for the TOF measurement is provided by the 
PIPS\textsuperscript{\textregistered} detector on which the source is built and the
``stop'' is provided by the CEM pulse.   In this case the 45~V bias of the 
PIPS\textsuperscript{\textregistered} detector is provided by a battery, so that the entire 
source can be floated to high voltage as required by the different configurations of the 
spectrometer.      Both signals from the source and the CEM are digitized with flash 
ADCs (FADCs) at 250~MS/s.    The signal from the source is used to trigger an acquisition.
While a fast preamplifier is used for the CEM pulse, the signal from the source is fed 
into an Ortec model 142 preamplifier followed by a model 474 timing filter amplifier 
(TFA) with 100~ns shaping.  Timing thresholds are applied offline.

\begin{figure}
\includegraphics[angle=90,scale=0.46]{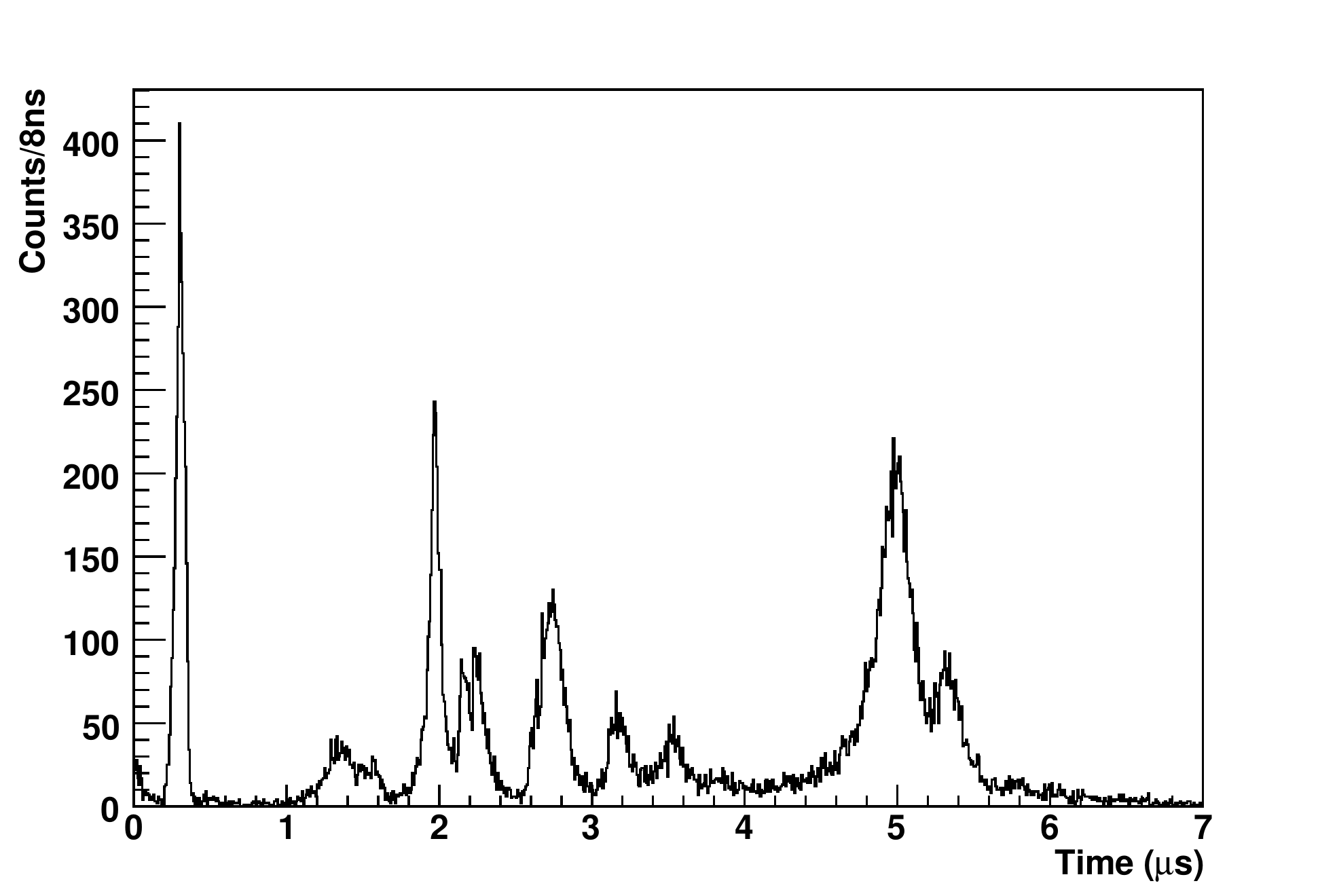}
\caption{\label{fig:Ba_tof} Time-of-flight spectrum for a typical Ba source.    The source 
has a measured $^{148}$Gd activity of 173~Bq and a 15~nm thick BaF$_2$ coating.   The Ba$^+$
yield, measured as the integral of the main peak, is of 2.7~Hz.}
\end{figure} 

A typical TOF spectrum obtained for the intermediate values of $E_{\rm source} = 550$~V/cm 
and $K = 2800$~eV is shown in Figure~\ref{fig:Ba_tof}.   The prominent peak around 
5~$\mu$s has a position and width consistent with those predicted by a SIMION~\cite{SIMION}
simulation for Ba$^+$ ions.    Such simulation includes the correct natural 
isotopic mix for the atomic mass of Ba and an initial energy distribution derived by the
SRIM output, as explained below.
A fit of this peak to five Gaussian functions representing the main Ba isotope masses
provides an estimate for the spectrometer FWHM resolution of 5.8~amu at $\simeq$137~amu.
The assignment of the peak to Ba$^+$ is corroborated by the observation 
that no such peak is present in a source where the BaF$_2$ coating is replaced by Al.     
In addition, and maybe most importantly, Ba deposited from a source of this type onto
a clean silicon substrate was then desorbed and resonantly ionized with two lasers.
From this very specific result it is clear that positive Ba ions are indeed emitted by 
the source.

According to SRIM the energy spectrum of the Ba emitted by the source peaks at zero and 
decreases monotonically, with 17\% of the spectrum having energies greater than 300~eV.
Since, as already mentioned, the SRIM simulation is unable to distinguish between neutral 
and charged states, a different method is required to estimate the expected ion fraction,  
$\alpha = n_+ / n_{\rm total}$.  Ideally this ratio would be known as a function of 
Ba energy.  Unfortunately, a general understanding of the ion fraction in sputtering 
is still an open question, even for pure metals, and data is completely lacking for ionic 
compounds, such as BaF$_2$~\cite{Wucher}.   Some models and experimental data describe $\alpha$
by a decaying exponential in the inverse velocity plus a constant term, with the transition 
taking place around a critical velocity of $\sim$2~cm$/\mu$s~\cite{Wucher}.  The net result 
of this function combined with the sputtering energy distribution is that the fraction of 
ions is low at low energy and approaches unity at higher energy.  
The transition takes place around 2~cm$/\mu$s, or $\sim$300~eV in the case of Ba.  

The actual energy distribution of Ba$^+$ ions affects the transport efficiency of the 
spectrometer significantly.  While such efficiency is calculated to be 85\% for the
entire energy distribution produced by SRIM, it is only 8\% for the part of such energy
distribution exceeding 300~eV.   Also the TOF peak position and width are to some 
extent predicted to vary for different regions of the emittance of the source.     
A simple consistency check for the Ba$^+$ yield can be made by assuming in the simulation 
that only ions (neutrals) are produced above (below) a certain energy cutoff.    The yield, 
TOF peak position and width are then compared with data for a variety of values of 
$E_{\rm source}$ and $K$.     A reasonably good description of the data occurs for 
a 300~eV cutoff.    The simulated rate reproduces the data within a factor of two when 
$E_{\rm source}$ and $K$ are varied from 550~V/cm and 2800~eV to the extremes mentioned above.   
With the same cutoff value the simulation also provides a good prediction for the Ba$^+$ 
TOF peak position (better than 10\%) and width (better than 25\%), again when scanning 
$E_{\rm source}$ and $K$ from their central values to the extremes.

For the spectrum in Figure~\ref{fig:Ba_tof} the Ba$^+$ yield is measured to be 2.7~Hz.    Using the 
$\alpha$ activity of 173~Bq, SRIM predicts a rate of Ba emission of 430~Hz for the 15~nm thick
BaF$_2$ coating (see Figure~\ref{fig:thickness}).    Assuming that the ions comprise only
those Ba atoms ejected with energy above 300~eV, or, according to SRIM 17\% of the total, 
the Ba$^+$ emission rate is calculated to be 73~Hz.  Using the spectrometer efficiency of 8\% 
appropriate for this energy range, a rate of 5.8~Hz is predicted to arrive at the CEM.  
The ratio between this figure and the measured rate of 2.7~Hz is consistent with typical CEM 
efficiencies for ions with the velocities expected here.

\begin{table}[h]
\begin{center}
	\caption{Possible assignment of the TOF peaks for the source in Fig.~\ref{fig:Ba_tof}.    
	The associations of H$^+$ and Ba$^+$ (in italics) are assumed and used to compute the 
	mass values for the other peaks.}
  \begin{tabular}{ |c |c |c| }
    \hline
    Time       & Measured mass & Possible assignment    \\ 
    ($\mu$s)   & (amu)				 & (known mass in [$\;$])     \\ \hline
    {\it 0.31} & {\it 1.01}    & {\it H$^+$ [1.01]}     \\ \hline
    1.44			 & 12.9          & C$^+$ [12.0]           \\ \hline
    1.99       & 23.5          & Na$^+$ [23.0]          \\ \hline
    2.16       & 27.6          & Si$^+$ [28.1]          \\ \hline
    2.23       & 29.3          & --                     \\ \hline 
    2.75       & 43.5          & C$_2$H$_5$O$^+$ [45.1] \\ \hline
    3.19       & 57.9          & C$_3$H$_8$O$^+$ [60.1] \\ \hline
    3.55       & 70.9          & Ba$^{2+}$ [68.7]       \\ \hline
    {\it 4.99} & {\it 137.3}   & {\it Ba$^+$ [137.3]}   \\ \hline
    5.33       & 156.2         & BaF$^+$ [156.3]        \\ \hline
  \end{tabular}
  \label{table:peaks}
\end{center}
\end{table}

The TOF spectrum for light masses is similar to the one observed for the source coated with Al.
Its interpretation nevertheless poses some challenges because of the limited resolution 
of the spectrometer, the large emittance of the source and the difficulty in predicting the 
charge to neutral ratio as a function of the species and initial energy.    A possible assignment,
obtained using as reference the Ba$^+$ peak and the peak at 0.31~$\mu$s, assuming to be due to 
H$^+$ is shown in Table~\ref{table:peaks}.    The presence of a prominent Na$^+$ peak is not
surprising, given the common presence of sodium contamination on surfaces.    Emission of some Si
from the substrate is not unexpected, while C$_2$H$_5$O$^+$ and C$_3$H$_8$O$^+$ are commonly 
observed in mass spectra of the IPA that is used in the Gd plating process. No compelling assignment 
is available for the peak at 2.23~$\mu$s.  Finally, the satellite peak at a mass larger than 
that of Ba is not present in the Al coated source and it may be attributed to BaF$^+$.

\section{Conclusions}

We have built a simple, inexpensive and compact source of single Ba$^+$ ions using the recoils
of an $\alpha$ emitting radionuclide to dislodge Ba atoms from a BaF$_2$ coating.   The source
uses a surface barrier detector as a substrate, so that a tag is provided for the ion emission.
Its simplicity and the possibility of delivering tagged ions in various environments make it 
attractive in applications where a low rate
of individual ions is desirable.      In addition the emission of neutral Ba, not useful for 
our purposes, may find other applications.   The replacement of the BaF$_2$ with other materials 
provides the means of emitting ions from a large number of different species.

\begin{acknowledgments}

We would like to thank K.~Moody (LLNL) and L.~Moretto (UC Berkeley and LBNL) for their 
radiochemistry advice. The support from the staff of the Stanford Health Physics group 
and, in particular, of D.~Menke, is also gratefully acknowledged.   We are also indebted 
to H.~Newman and R-Y.~Zhu (Caltech) for providing a high quality crystal of barium fluoride.   
Finally, we would like to thank Canberra Semiconductor NV and, in particular, M.~Morelle 
for advice and for providing us with un-encapsulated samples of their detectors used for 
practicing the different processing steps.   
This work was supported in part by NSF grant PHY-0652416.

\end{acknowledgments}


\end{document}